\documentclass[preprint,preprintnumbers,amsmath,amssymb]{revtex4}
\usepackage{graphicx}
\usepackage{dcolumn}
\usepackage{bm}
\begin{document}
\title{Morgan-Morgan-NUT disk space via the Ehlers transformation }
\author{D.~Momeni$^{a}$\footnote{~dmomeni@phymail.ut.ac.ir}, M.~Nouri-Zonoz$^{a,b,c}$\footnote{~nouri@khayam.ut.ac.ir} and R. Ramazani-Arani
$^{a}$\footnote{~ramazani@phymail.ut.ac.ir}}
\address{$^{a} $ Department of Physics, Universiy of Tehran, 
North Karegar St., 14395-547, Tehran, Iran
\\$^{b}$ Institute for studies in theoretical physics and
Mathematics, P O Box 19395-5531, Tehran, Iran\\
$^{c}$ Institute of Astronomy, Madingley Road, Cambridge, CB3 0HA, UK}
\begin{abstract}
Using the Ehlers transformation 
along with the gravitoelectromagnetic approach to stationary spacetimes we start
from the Morgan-Morgan disk spacetime (without radial pressure) as the seed metric
and find its corresponding stationary spacetime. As expected from the Ehlers 
transformation the stationary spacetime obtained suffers from a NUT-type singularity
and the new parameter introduced in the stationary case could be interpreted as the
gravitomagnetic monopole charge (or the NUT factor). As a consequence of this singularity 
there are closed timelike curves (CTCs) in the singular region of the spacetime. 
Some of the properties of this spacetime including its particle velocity distribution, 
gravitational redshift, stability and energy conditions are discussed.  
\end{abstract}
\pacs{PACS No.}
\maketitle
%%%%%%%%%%%%%%%%%%%%%%%%%%%%%%%%%%%%%
\section{Introduction}
Finding a new solution of Einstein field equations from a given
solution of the same field equations has a long history and
mathematically is related to the problem of finding a solution
to a set of  nonlinear partial differential equations from another
known solution. Here we make use of a very special transformation
introduced by Ehlers [1] through which one could start from a
static spacetime and end up with a stationary solution. 
We apply this transformation to two static spacetimes, namely 
Schwarzschild metric and Morgan-Morgan disk space [2]
and in the meantime employ the concepts introduced
in the gravitoelectromagnetic approach (1+3 formalism) to
spacetime decomposition to yield a physical interpretation 
of the spacetime found through the transformation. The use 
of the gravitoelectromagnetic concepts are justified by the fact
that the main equation of the Ehlers transformation
includes the gravitomagnetic potential of the stationary spacetime.
The outline of the paper is as follows. In section II we
introduce briefly gravitoelectromagnetism and the Ehlers transformation 
and use the latter to derive, as an example, the NUT solution [3] from 
Schwarzschild spacetime and also Chazy-Curzon-NUT (CC-NUT) solution from 
Chazy-Curzom static metric [4,5]. 
In section III we introduce briefly the static Morgan-Morgan disk space 
(without radial pressure) and some of its properties. In section IV using Ehlers 
transformation we find the corresponding stationary spacetime representing the spacetime 
of a finite disk of mass $M$ and gravitomagnetic charge $l$. In section V we discuss 
some of the properties of this stationary disk including its particle velocity 
distribution, gravitational redshift, stability and energy conditions. 
In the last section we summarize the results obtained.
%%%%%%%%%%%%%%%%%%%%%%%%%%%%%%%%%%%%%%%%%%%%%%%%%%%%%%%%%%%%%%%%%%%%%%%%%%%%
\section{Gravitoelectromagnetism and Ehlers transformation }
The $(1+3)$-decomposition (threading) of a spacetime  by a
congruence of timelike curves (observer worldlines) leads to the
following splitting of the spacetime interval element [6];
$$ds^{2}=dT^{2}-dL^{2},\eqno(1)$$
where $dL$ and $dT$ are defined to be {\it the invariant spatial
and temporal length elements} of two nearby events respectively.
They are constructed from the normalized tangent vector $u^{a}=
{\xi^a \over |\xi|}$  to the timelike curves in the following way
\footnote{Note that Latin indices run from 0 to 3 while the Greek
ones from 1 to 3 and throughout we use gravitational units where
c=G=1.};
$$dL^{2}= h_{ab}dx^{a}dx^{b},\eqno(2)$$
$$dT= u_{a}dx^{a},\eqno(3)$$
where
$$h_{ab}= -g_{ab}+u_{a}u_{b},\eqno(4)$$
is called the {\it projection tensor}. Taking $h \equiv |\xi|^2$
and $A_a  =\equiv -{\xi_a \over |\xi|^2} $, equations (1) and (4)
can be written in the following alternative forms;
$$ds^{2}=  h(A_a {\rm d}x^a)^2 - h_{ab}{\rm d}x^a
{\rm d}x^b\;\;\; ; \;\;\; h_{ab}= -g_{ab}+h A_{a}A_{b}.\eqno(5)$$
Using the {\it preferred coordinate} system in which the timelike curves
are parameterized by the coordinate time $x^{0}$ of the comoving
observers;
$$\xi^a = (1,0,0,0)\;\;\;\;\;\ ; \;\;\;\;
A_a = (-1, - {g_{0\alpha}\over g_{00}}),\eqno(6)$$ and the above
spatial and spacetime distance elements will take the following
forms [6,7];
$$dL^2 \equiv dl^{2}=\gamma_{\alpha\beta}dx^{\alpha}dx^{\beta},\eqno(7)$$
$$ds^{2}=e^{2U}(dx^{0}-A_{\alpha}dx^{\alpha})^{2}-dl^{2},\eqno(8)$$
where
$$e^{2U}\equiv g_{00} \;\;\;\;\; ,\;\;\;\;\; A_{\alpha} = -g_{0\alpha}/g_{00},\eqno(9)$$
and
$$\gamma_{\alpha\beta} = (-g_{\alpha\beta} +{g_{0\alpha} g_{0\beta}\over g_{00}}).\eqno(10)$$
Introducing $U$ and $A$ as the {\it gravitoelectric} 
and {\it gravitomagnetic} potentials respectively, their corresponding fields are
\footnote{For reviews on the subject of Gravitoelectromagnetism see references [8-10].} [7];
$${{\bf E}_g} = -\nabla {U},\eqno(11)$$
$${\bf B}_g = {\rm curl} {\bf A },\eqno(12)$$
where the differential operators are defined in the $\gamma$ space.
On the other hand Ehlers transformation, also called {\it gravitational duality
rotation}, in its original form states that if
$$g_{mn}dx^m dx^n= e^{2U}(dx^0)^2 -e^{-2U}d{\tilde l}^2,\eqno(13)$$
with $d{\tilde l}^2 = e^{2U} dl^2$  representing the {\it conformal} spatial distance,
is the metric of a static exterior spacetime, then,
$${\bar g}_{mn}dx^m dx^n = (\alpha {\rm cosh}(2U))^{-1} (dx^0 - A_\beta dx^\beta)^2
- \alpha {\rm cosh}(2U)d{\tilde l}^2,\eqno(14)$$
(with $\alpha = constant > 0$, $U=U(x^\alpha)$ and $A_\beta =
A_\beta(x^\alpha)$) would be the metric of a stationary exterior spacetime
provided that $A_\alpha$ satisfies the following equation;
$$\alpha\sqrt{\tilde \gamma} \varepsilon_{\alpha\beta\eta} U^{,\eta} =
A_{[\alpha,\beta]},\eqno(15)$$
where $\tilde \gamma = {\rm det}{\tilde \gamma}_{\mu\nu}$, ${\tilde \gamma}_{\mu\nu}$ 
being the conformal spatial metric.
Hereafter we call the above equation the Ehlers equation.\\
Considering the Ehlers
transformation and in particular Ehlers equation (15) in the 
context of gravitomagnetism, then finding a
stationary solution corresponding to a given static solution,
amounts to finding the gravitomagnetic potential $A_\alpha$
related to a given static (potential) function $U$.
In what follows, as an example of the procedure, we derive the NUT
space [8] as the stationary spacetime with a radial gravitomagnetic field
which is the gravitational dual of Schwarzschild space.
%%%%%%%%%%%%%%%%%%%%%%%%%%%%%%%%%%%%%%%
\subsection{NUT space from Schwarzschild through Ehlers transformation}
Starting from the schwarzschild mertic in the form (13) in which ;
$$U_c= {1\over 2}\ln(1-2m/r)+ C,\eqno(16)$$
and \footnote {Note that we have added a constant to the potential using the fact that solutions with potentials differing in additive constants are equivalent.}
$$d{\tilde l}^2 = dr^2 +r^2 \exp(2U)(d\theta^2 + \sin^2\theta d\phi^2),\eqno(17)$$
we find that in this case the Ehlers equation (15) reduces to,
$$\alpha r^2 e^{2U} \sin \theta {\tilde \gamma}^{rr}U_{,r} = -{1\over 2}A_{\phi,\theta},\eqno(18)$$
in which we set $A_{\theta,\phi}=0$ to keep the resulted spacetime single valued
and axially symmetric. Using (16) the gravitomagnetic field is found to be,
$$A_\phi(\theta) = 2\alpha m \cos\theta. \eqno(19)$$
So the resulted stationary metric is given by (14) as;
$$ds^2 = {r(r-2m)\over \alpha f(r)}\left( dt-2\alpha m \cos\theta d\phi\right)^2 - {\alpha f(r)\over r(r-2m)}dr^2 -\alpha f(r) d \Omega ^2,\eqno(20)$$ 
where 
$$f(r) =  r^2({1+c_1^2\over 2c_1}) + 2m^2c_1 - 2mrc_1 \;\;\;\;\ , \;\;\;\;\ c_1 = e^{2C}. \eqno(21) $$ 
This is the well known NUT space [3,11]. To recover the Schwarzschild spacetime metric as $\alpha \rightarrow 0$ we need
$$  \alpha={2c_1 \over 1+c_1^2}\;\;\; , \;\;\; |c_1| \leq 1. \eqno(22)$$
Now applying the following changes of variables and redefinitions of constants ;
$$ M=m({1-c_1^2\over 1+c_1^2}) \;\;\; , \;\;\; \alpha m = {\it l}\;\;\; , \;\;\;(r- l c_1) = R, \eqno(23)$$
to the line element (20) we have,
$$ds^2 = {R^2-2MR-l^2\over R^2 + l^2}(dt-2l\cos \theta d\phi)^2 - {R^2+l^2\over R^2 - 2MR -l^2}dR^2
- (R^2 + l^2)d \Omega ^2,\eqno(24)$$
which is the NUT space in its more common form [8,11,12]. Note that having started from a potential with $C=0$ (or $c_1=1$) we would have ended up with the {\it pure }($M=0$) NUT space. In other words although adding the constant $C$ in the static case does not correspond to a different spacetime, in the Ehlers transformation it does. This actually answers another question of concern, namely, what is the seed metric (in Ehlers transformation) of pure NUT space?. By the above argument for both NUT and pure NUT spaces the seed metic is the Schwarzschild metric.
%%%%%%%%%%%%%%%%%%%%%%%%%%%%%%%%%%%%%
\subsection{stationary spacetimes with $A_\alpha = A_\alpha (r,\theta)$}
As another example of the application of the Ehlers transforemation in this section we 
consider a family of solutions for a gravitomagnetic potential of the form
$${\bf A} = A_{\phi} (r, \theta){\bf {\hat\phi}},\eqno(25)$$
(In the spherical or the schwarzschild-type coordinates
 $(t,r,\theta,\phi)$ ) which according to the $1+3$ formalism
[7-9] could, in principle, produce stationary spacetimes with 
gravitomagnetic field components
$B_g(r,\theta){\bf {\hat r}}$ and $B_g(r,\theta){\bf {\hat \theta}}$.
Using Ehlers transformation one could easily show that for an $\bf A$
of the type (25) one ends up with the equation $\nabla^2U=0$
for $U$ with the general solution [11]
$$U = \Sigma_{n=0}  a_n r^{-(n+1)} P_n(cos\theta). \eqno(26)$$
For the simplest case, i.e. $n=0$ we have $U\propto{1 \over r}$
which, in the static case corresponds to the 
Chazy-Curzon metric [4,5] and in the stationary case 
(through Ehlers transformation) to a spactime dual to Chazy-Curzon with 
a gravitomagnetic potential of the form $A_\phi(\theta) =
a cos\theta + b$ [13,14]. It is notable that this has exactly the same form as 
the gravitomagnetic potential of NUT space and consequently the parameter 
$\alpha$ could be interpreted as the NUT factor (or magnetic mass). Another
example in which Ehlers transformation is examined on a static spacetime to 
yield an stationary space is the case of {\it cylindrical} NUT space [15] where
it is shown that it could be obtained from Levi-Civita's static cylindrically 
symmetric solution.
%%%%%%%%%%%%%%%%%%%%%%%%%%%%%%%%%%%%%%%%%%%%%%%%%%%%%%%%%%%%%%%%%%%%%%%%%%%%%%%
\section{static spacetime of a pressureless (Morgan-Morgan) disk}
Static axially symmetric spacetimes in the Weyl canonical
coordinates are given by the following general form [11];
$$ds^2 = e^{-2U}\left[ e^{2k} (d\rho^2 + dz^2) + \rho^2
d\phi^2\right] - e^{2U}dt^2, \eqno(33)$$
in which $U$ and $k$  are functions of $\rho$ and $z$.
Morgan and Morgan [2,11] found the following
solution for a finite disk (of mass $M$ and radius
$a$) without radial pressure in terms of the oblate ellipsoidal coordinates $(\xi , \eta)$;
$$U=-{M\over a}\left( arccot \xi + {1\over 4}[(3\xi^2+1)arccot \xi-3\xi]
(3\eta^2-1)\right),\eqno(34)$$
and
$$k={9\over 4}M^2\rho^2 a^{-4}\left[ ({\rho \over a})^2 B^2(\xi) -
(1+\eta^2)A^2(\xi) - 2\xi(1-\eta^2)A(\xi) B(\xi)\right],\eqno(35)$$
where
$$A(\xi) = \xi {\rm arccot} \xi -1 \;\;\; , \;\;\; B(\xi) = {1\over 2}[{\xi \over
1+\xi^2} - {\rm arccot}\xi],\eqno(36)$$
in which the connection between the ellipsoidal $(\xi, \eta )$ and
Weyl $(\rho, z)$ coordinates;
$$ \rho^2 = a^2(1+\xi^2)(1-\eta^2)\;\;\; , \;\;\; z=a\xi\eta \; , \;
|\eta|\leq 1 \;\;\; , \;\;\; 0\leq\xi\leq \infty, \eqno(37)$$
infer that the disk is located at $\xi =0 \; , \;|\eta|\leq 1 $ .
In [16] the authors have found the form of the same metric in
Weyl canonical coordinates as the imaginary part of a complexified
bar with the functions $U$ and $k$ as follows;
$$U = -{M\over 2ia} ln|{Re[R] - ia\over Re[R] + ia}|,\eqno(38)$$
$$k = -{1\over 2}({M\over 2})^2 ln|{Re[R]^2 + a^2\over R^2}|.\eqno(39)$$
where $R = \sqrt{\rho^2 + (z-ia)^2}$.
Exact thin disk solutions of Einstein field equations are discussed extensively in different 
contexts in the literature. The above static solution, first derived by Morgan and Morgan [2], is considered to represent the 
spacetime of a finite disk of counterrotating particles.
Properties of static counterrotating disks have been studied in [17,18]. Superposition of a central 
black hole with a surrounding thin disk of counterrotating particles is discussed in [19] and 
infinite self-similar counterrotating disks were studied in [20,21].
In the following section using the Ehlers transformation we obtain the stationary spacetime
corresponding to the Morgan-Morgan static disk space. To our knowledge the exact NUT extension 
of the above Morgan-Morgan disk space has not appeared in the literature.
%%%%%%%%%%%%%%%%%%%%%%%%%%%%%%%%%%%%%%%%%%%%%%%%%%%%%%%%%%%%%%%%%%%%%
\section{stationary spacetime of a finite NUTTY disk}
Starting with the Morgan-Morgan static disk space given by equations (34-36) and
using the fact that $3\eta^2-1 = 2 P_2(\eta)$ we write (34) as
follows
$$U_c=-{M\over a}\left( {\rm arccot} \xi + {1\over 2}[(3\xi^2+1){\rm arccot} \xi-3\xi]
P_2(\eta)\right)+C.\eqno(40)$$
Now choosing an axially symmetric gravitomagnetic potential ${\bf
A}=A_\phi(\xi,\eta){\hat{\bf\phi}}$, by Ehlers transformation we have;
$$A_{\phi , \eta} = 2\alpha a (1+\xi^2)U_{,\xi},\eqno(41)$$
$$A_{\phi , \xi} = -2\alpha a (1-\eta^2)U_{,\eta},\eqno(42)$$
where $\alpha$ is the Ehlers transformation parameter (duality
rotation parameter) and $a$ is the radius of the disk with mass
$M$. Since the right hand sides of the above two equations include Legendre
polynomials $P_0(\eta)$ and $P_2(\eta)$ we can expand the
gravitomagnetic potential as follows;
$$ A_\phi (r,\theta) = f_1 (\eta) P_1 (\xi) + f_3 (\eta) P_3
(\xi).\eqno(43)$$
By substituting the above potential into the equations (41) and
(42) we end up with the following solution for the gravitomagnetic potential;
\begin{eqnarray*}
\lefteqn{A_\phi(\xi,\eta) = {3l}\xi \eta(1-\eta^2)(1+\xi^2)\arctan \xi+} \hspace{1.25in} \\ 
 & & {3\over 2} \eta l \left[(\pi \xi^3 -2\xi^2 -4/3 + \pi\xi)\eta^2 - 
 (\pi \xi^2 -2\xi + \pi)\xi \right] + C_1, \hspace{.75in} (44)
\end{eqnarray*}
where as before $l=M\alpha$ is the NUT factor. To regain the Morgan-Morgan space 
as $l\rightarrow 0$, we choose $\alpha = 2c_1$ where $c_1=e^{2C}$, 
so that the line element of the stationary spacetime of a finite 
disk with mass M and magnetic mass (NUT factor) $l$ is given by ;
\begin{eqnarray*}
ds^2 &= &({l^2\over 4m^2}e^{2U}+e^{-2U})^{-1}\left[ dt - A_\phi (\xi,\eta)
d \phi \right]^2  \\
     &-& 
a^2({l^2\over 4m^2}e^{2U}+e^{-2U})\left(e^{2k} (\xi^2 +
\eta^2)({d\xi^2\over 1+\xi^2} + {d\eta^2\over 1-\eta^2}) +
(1+\xi^2)(1-\eta^2)d\phi^2 \right).\hspace{.4in}(45)
\end{eqnarray*}
This is verified to be an exact solution to the vacuum Einstein field equations 
using the Maple tensor package \footnote{We used Maple 8, 
Waterloo Maple Inc., 2002.}. As the original Morgan-Morgan disk space is asymptotically flat,
it is not difficult to see that the above stationary spacetime shares the same property.
Hereafter we call the above solution the MM-NUT disk space. 
The presence of the Dirac-type (NUT-type) singularity could 
be inferred from the form of the gravitomagnetic potential (44), as one could not make it vanish 
simultaneously for both the positive ($z> 0 \equiv \eta =  1$) and negative ($z> 0 \equiv \eta = - 1$) half-axes ($\rho = 0$) by any choice of the constant $C_1$. In other words only half of the axis could be made regular, either the positive half-axis (for $C_1= 2l$) or the negative half-axis (for $C_1= -2l$). Therefore a shrinking loop around the singular half-axis will have a non-zero circumference even when $\rho \rightarrow 0$. Now consider a $t=Constant$ hypersurface and take the lower half-axis ($\eta = -1$) to be singular ( i.e. $C_1= 2l$). In this case the coefficient of $d\phi^2$ in (45) becomes positive at sufficiently small $\rho$ ($\eta \approx -1$); hence $\phi$ becomes a timelike coordinate and being a cyclic variable with a period of $2\pi$, this means that the spacetime contains {\it closed timelike curves} in the singular section; another characteristic of NUT-type singularity.

%%%%%%%%%%%%%%%%%%%%%%%%%%%%%%%%%%%
\section{Particle velocity, stability, energy conditions and gravitational redshift }
In this section we calculate particle velocity distribution of the MM-NUT
(stationary) disk found in the previous section. We also discuss
its stability and energy conditions and obtain the gravitational 
redshift suffered by the photons emitted from different radii on the disk. 
Hereafter we take a disk with unit radius i.e. we set $a=1$ for convenience.
%%%%%%%%%%%%%%%%%%%%%%%%%%%%%%%%%%%%%%%%%%%%
\subsection{Particle velocity}
The general form of the metric for stationary axially symmetric spacetimes
(in cylindrical coordinates) is given by;
$$ ds^2 = g_{tt}(\rho,z)dt^2 + 2g_{t\phi}(\rho,z)dtd\phi + g_{\alpha\alpha}(\rho,z)(dx^\alpha)^2 \;\;\;\ \alpha = \rho,z,\phi. \eqno(46)$$
For a zero angular momentum observer (ZAMO) in these spacetimes, the velocity of a particle in a circular motion (in the $\rho=Constant, z=0 $ plane), is given by [22];
$$v^2_{\phi}={g_{\phi\phi}^2 \over g_{t\phi}^2 -  g_{tt}g_{\phi\phi}}\left( {d\phi\over dt} + {g_{t\phi}\over g_{\phi\phi}} \right)^2.\eqno(47)$$
Therefore we need to find $d\phi \over dt$, for which, we use the energy-momentum 
tensor of the particle distribution which in turn could be calculated by the formalism
of distributions [19,23]. Using the general form of the energy-momentum tensor of a system 
of particles [7],
$$T^{\mu\nu} = \Sigma m_i(-g)^{-{1\over 2}} {dx^\mu \over dt}{dx^\nu \over ds}\delta(x^i - x),\eqno(48)$$
and the fact that ${d\phi \over dt}=({T^{\phi\phi}\over T^{tt}})^{1/2}$ we end up with
the folowing final result ;
$$({d\phi \over dt})^2 = {T^{\phi\phi}\over T^{tt}} = {b_{\rho\rho}(g^{\rho\rho}g^{\phi\phi}) + b_{tt}(g^{tt}g^{\phi\phi}- g^{t\phi}g^{t\phi})\over b_{\rho\rho}(g^{tt}g^{\rho\rho}) + b_{\phi\phi}(g^{tt}g^{\phi\phi}- g^{t\phi}g^{t\phi})}, \eqno(49)$$
in which, through the Einstein field equations, the components of the energy-momentum tensor are given by
\footnote{For an axially symmetric metric the non-zero components of this tensor (at $z=0$ plane) 
are given by
$$T^\phi_\phi = {g^{zz}\over 2}(g^{\rho\rho}b_{\rho\rho} + g^{\phi\phi}b_{\phi\phi} + g^{t\phi}b_{t\phi})\; , \;
T^t_t = {g^{zz}\over 2}(g^{\rho\rho}b_{\rho\rho} + g^{tt}b_{tt} + g^{t\phi}b_{t\phi}) \; , \; 
T^t_\phi = -{g^{zz}\over 2}(g^{tt}b_{t\phi} + g^{t\phi}b_{\phi\phi})$$},
$$T^a_b = {1\over 2}\left[b^{az}\delta^z_b - b^{zz}\delta^a_b + g^{az}\delta^z_b -g^{zz}b^a_b + b_c^c(g^{zz}\delta^a_b - g^{az}\delta^z_b)\right]\delta(z),\eqno(50)$$
where $b_{ab}$ are the components of the discontinuity in the first derivative of the 
metric tensor (at $z=0$),
$$b_{ab} = g_{ab,z}|_{z=0^+} - g_{ab,z}|_{z=0^-} = 2g_{ab,z}|_{z=0^+}.\eqno(51)$$
Computing these elements for the metric (45) we have;
$$b_{tt}={30M\sqrt{1-\rho^2}F_{-}(\rho)\over {F_{+}(\rho)}^2},\eqno(52)$$
$$b_{\rho\rho}=2e^{{9\over 2}M^2\rho^2({1\over 16}\pi^2\rho^2 - 2 +  \rho^2)}\left(15M\sqrt{1-\rho^2}F_{-}(\rho)
+{45\over 2}\pi M^2\rho^2\sqrt{1-\rho^2}F_{+}(\rho)\right),\eqno(53)$$
$$b_{\phi\phi}=30M\sqrt{1-\rho^2}F_{-}(\rho)\left(\rho^2 + {G(\rho)\over F_{+}^2(\rho)}\right),\eqno(54)$$
where 
$$F_{\pm}(\rho) = {l^2\over 4M^2}e^{{3\over4}M\pi(\rho^2-2)}\pm e^{-{3\over4}M\pi(\rho^2-2)},\eqno(55)$$
and
$$G(\rho)\equiv A^2_{\phi}(z=0)=(2l-2l[1-\rho^2]^{3\over 2})^2. \eqno(56)$$
Now substituting (52-56) into (50) and use the outcome in (49) and (47) we obtain the 
follwoing expression for the particle velocity,
$$v_{\phi}={1\over {\rho F_{+}(\rho)}} \left( F_{+}^2(\rho)\rho^2-G(\rho) \right)\left(\Omega(\rho)+{G^{1/2}(\rho)\over F_{+}^2(\rho)\rho^2-G(\rho)}\right), \eqno(57)$$
where 
\begin{eqnarray*}
\Omega(\rho)& = & 4\sqrt{3\pi}\left(3\pi{l^4\over M^4}\rho^2 \exp({3M\pi(\rho^2-2)\over 2})  
+ 4{l^4\over M^5}\exp({3M\pi(\rho^2-2)\over 2})\right.  \\
            & + &   
\left. 24\pi {l^2\over M^2}\rho^2 - 192 l \pi G^{1\over 2}(\rho)+576\pi l^2 \rho^2 (1- \rho^2) + 192\pi l^2 \rho^6 \right.  \\
            & + &
\left. 48\pi\rho^2\exp(-{3M\pi(\rho^2-2)\over 2})
-{64\over M}\exp(-{3M\pi(\rho^2-2)\over 2})\right)^{-{1\over 2}}.\hspace{1.4in} (58) 
\end{eqnarray*}
For the disk to be physical we impose the condition that the velocity of the particles at the rim of the disk do not exceed that of the light i.e.,
$$v_{\phi}(\rho=1)\leq 1. \eqno(59)$$
Imposing the above constraint in its extreme form (i.e. with the equal sign) on (57), we find the following relation between $l$ and $M$ 
$$l_{\pm}={M\over 3\pi M + 2}\exp({3\pi M \over 4})\left(-12\pi M^2 \pm 2\sqrt{36\pi^2M^4 -9\pi^2M^2 +4}\right), \eqno(60)$$
and consequently the following constraint on the maximum value of the disk mass,
$$ M_{max}^{MM-NUT} = {\sqrt{6}\over 12\pi}\left( 3\pi^2 - \sqrt{9\pi^4 - 
64\pi^2}\right)^{1\over 2}\approx 0.2427,\eqno(61)$$
for $l=-0.2204$.
This is a higher maximum value for the disk mass compared to that obtained by Morgan and Morgan [2],
$$M_{max}^{MM} = {2\over 3\pi}\approx 0.2122.\eqno(62)$$
In other words, addition of the NUT factor as an extra parameter, not only changes the static character of the spacetime (to stationary) but also allows for a higher mass of the disk to be acceptable physically. Note that this maximum value of mass is obtained for a negative value of $l$ which, by the form of the metric (45), is acceptable. In Fig. 1 the diagram of $l$ in terms of of $M$, given by the constraint (60), is shown and only those pair of values $(l,M)$ inside the loop formed by the $l_+$  and $l_-$ curves are acceptable physically. 
For a fixed disk mass $M$ the velocity of particles as a function of radius is shown in Fig. 2 for different values of $l$.

\begin{figure}
\begin{center}
\includegraphics[angle =-90,scale=0.7]{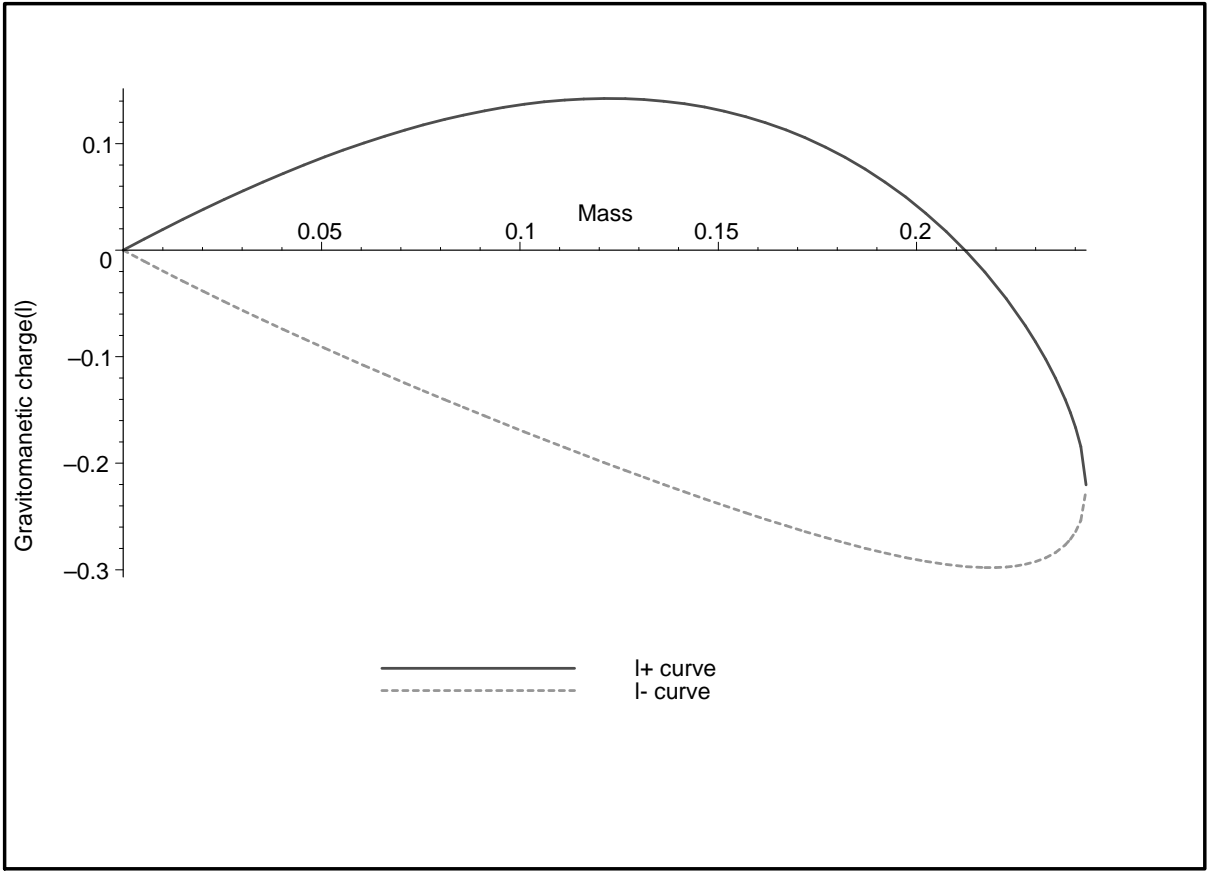}
\caption{$l$ as a function of $M$ by velocity constraint (60). The allowed values of $l$ and $M$ lie inside the loop formed by the two curves.}
\end{center}
\end{figure}

\begin{figure}
\begin{center}
\includegraphics[angle =-90,scale=0.6]{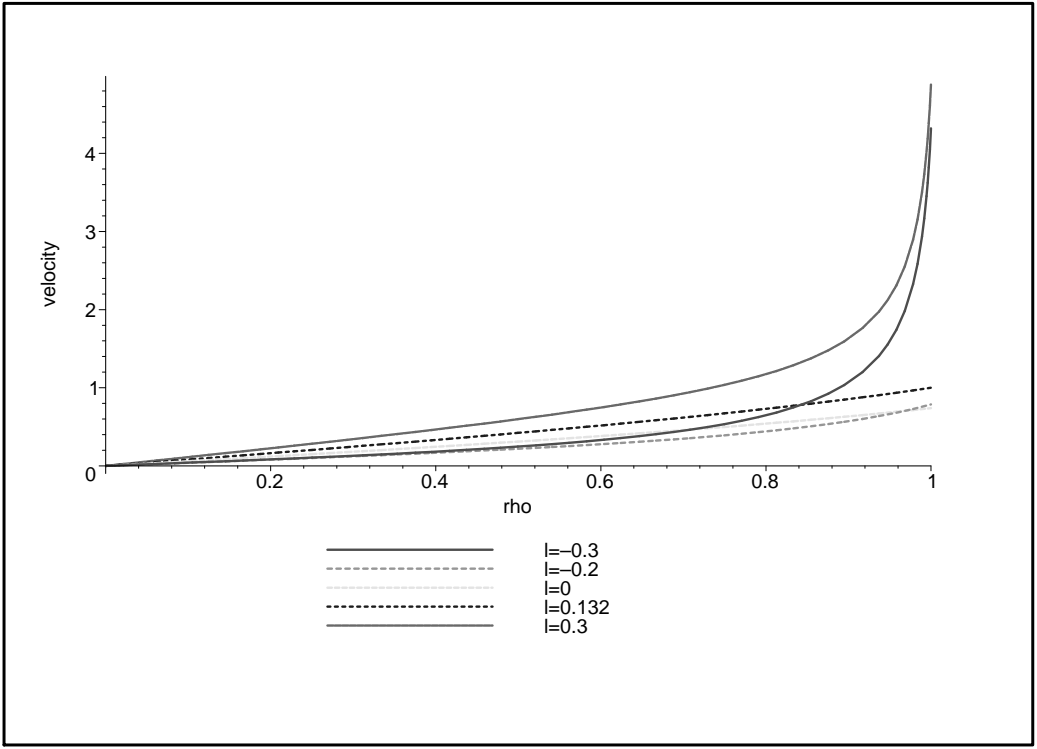}
\caption{Velocity of disk particles vs radius for a fixed mass $M=0.15$ and different $l$ values
including $l=0$ i.e. Morgan-Morgan disk. For two forbidden values of $l$ the velocity diverges 
and exceeds that of light.}
\end{center}
\end{figure}
%%%%%%%%%%%%%%%%%%%%%%%%%%%%%%%%%%%%%%%%%%%%%%%%%%%%
\subsection{Stability and energy conditions}
Having discussed the velocity distribution of the disk particles one could also discuss stability of particle
orbits under radial perturbations. To do so we use the extended Rayleigh criterion of stability under radial 
perturbations given by [23,24],
$${\rm {d L^2\over  d\rho}} > 0. \eqno(63)$$
where $L$ is the angular momentum of disk particles.
For circular geodesics in $z=0$ plane the specific angular momentum is given by;
$${\rm L} = g_{\phi\phi}\dot{\phi}+ g_{t\phi}\dot{t}= {1\over \sqrt{g_{tt}}}(g_{t\phi} 
+ g_{\phi\phi}{{\rm d}\phi \over {\rm d}t}).\eqno(64)$$
Substituting for ${\rm d}\phi \over {\rm d}t$ from (49) it could be seen that $\rm L^2$ is an increasing function of $\rho$ and satisfies the above criterion leading to the stability of circular orbits (Fig. 3). 
\begin{figure}
\begin{center}
\includegraphics[angle =-90,scale=0.6]{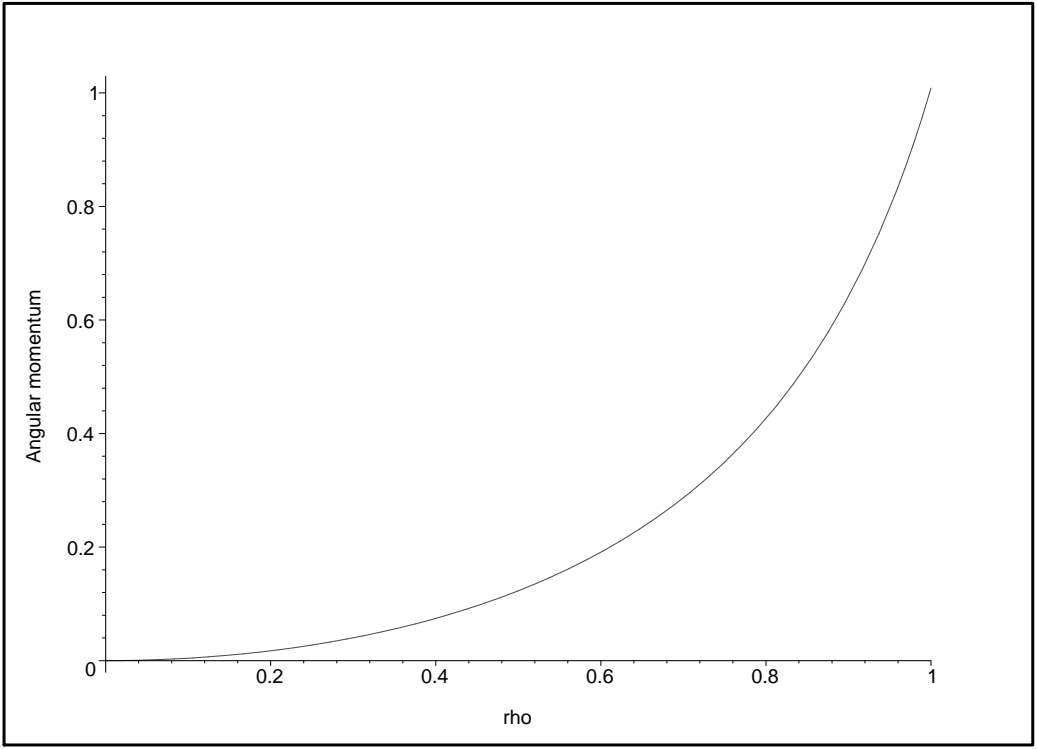}
\caption{Specific angular momentum as a function of $\rho$.}
\end{center}
\end{figure}
Since we have already confined the domain of the allowed values of $l$ and $M$ to those respecting the velocity of
light barrier, it is expected that the same domain of values respect the energy conditions as well. This is indeed 
the case for strong and weak energy conditions as we will see below but not for the dominant energy condition. To consider the energy conditions for MM-NUT disk space
we diagonalize the stress-energy tensor to find the principal 
pressures [23]. The Eigenvalue equation for the 
stress-energy tensor ;
$$T^a_b \xi^b= \lambda \xi^a \;\;\;\; a,b = t,\phi. \eqno(65)$$
constitutes the following eigenvalues
$$\lambda_{\pm} = {1\over 2}(T\pm \sqrt{D}),\eqno(66)$$
where 
$$D=(T^t_t - T^\phi_\phi)^2 + 4T^t_\phi T^\phi_t \;\;\;\;{\rm and}\;\;\;\; 
T=T^t_t + T^\phi_\phi. \eqno(67)$$
It could be easily checked that for the discriminant $D>0$ the eigenvalues $\lambda_{\pm}$ 
are the azimuthal pressure $P$ and energy density $\epsilon$ respectively [23].
In our case, using the non-zero components of the energy-momentum tensor (50), 
the discriminant,
$$D = 14400(1-\rho^2)M^6 e^{(-{9M^2\rho^2(\pi^2\rho^2 - 32 + 16 \rho^2)\over 16})}
{[l^2 e^{({3\pi M \over 4}(2-\rho^2))} - 4M^2 e^{({3\pi M \over 4}(2-\rho^2))}]^2\over
[l^2 e^{({3\pi M \over 4}(2-\rho^2))} + 4M^2 e^{({3\pi M \over 4}(2-\rho^2))}]^4}. \eqno(68)$$ 
is positive definite and the strong energy condition (SEC), 
$${\rm SEC} \equiv \epsilon + P =\lambda_{+} -\lambda_{-}=|D|^{1\over 2}\geq 0. \eqno(69)$$
is always satisfied. On the other hand the weak energy condition (WEC) and the dominant energy condition (DEC) could be written equivalently as the following conditions on the magnitude of the gravitomagnetic charge $l$,
$$ {\rm WEC}\; \equiv \; \epsilon = -\lambda_- \geq 0 \Longrightarrow |l| < 2M \sqrt{{4-3\pi M\rho^2 \over 4+3\pi M\rho^2}} e^{{3\pi M \over 4}(2-\rho^2)},\eqno(70)$$
$$ {\rm DEC}\; \equiv \; \epsilon \geq |P| \Longrightarrow |l| < 2M \sqrt{{2-3\pi M\rho^2 \over 2+3\pi M\rho^2}}e^{{3\pi M \over 4}(2-\rho^2)}.\eqno(71)$$
It could be seen that the most 
restricted situation happens at the rim of the disk, so, to ensure the above conditions
all over the disk we apply them at $\rho = 1$ i.e.,
$${\rm WEC} \equiv |l| \leq 2M \sqrt{{4-3\pi M \over 4+3\pi M}}
\exp({3\pi M \over 4}),\eqno(72)$$
$${\rm DEC} \equiv |l| \leq 2M \sqrt{{2-3\pi M \over 2+3\pi M}}
\exp({3\pi M \over 4}).\eqno(73)$$
In Fig. 4 the DEC, WEC and velocity conditions are shown in a diagram 
for the same range of values of $l$ and $M$ as in Fig. 1. It is noted that the WEC is already satisfied by those values of $l$ and $M$ respecting the velocity condition, but DEC could be violated for some values of $l$ and $M$ already allowed by the velocity condition and vice versa. We also note from (73) that, for the static Morgan-Morgan disk ($l=0$), DEC is equivalent to the velocity condition, in other words, the maximum allowed mass is given by $M_{max}= {2\over 3\pi}$ as in (62).
\begin{figure}
\begin{center}
\includegraphics[angle =-90,scale=0.6]{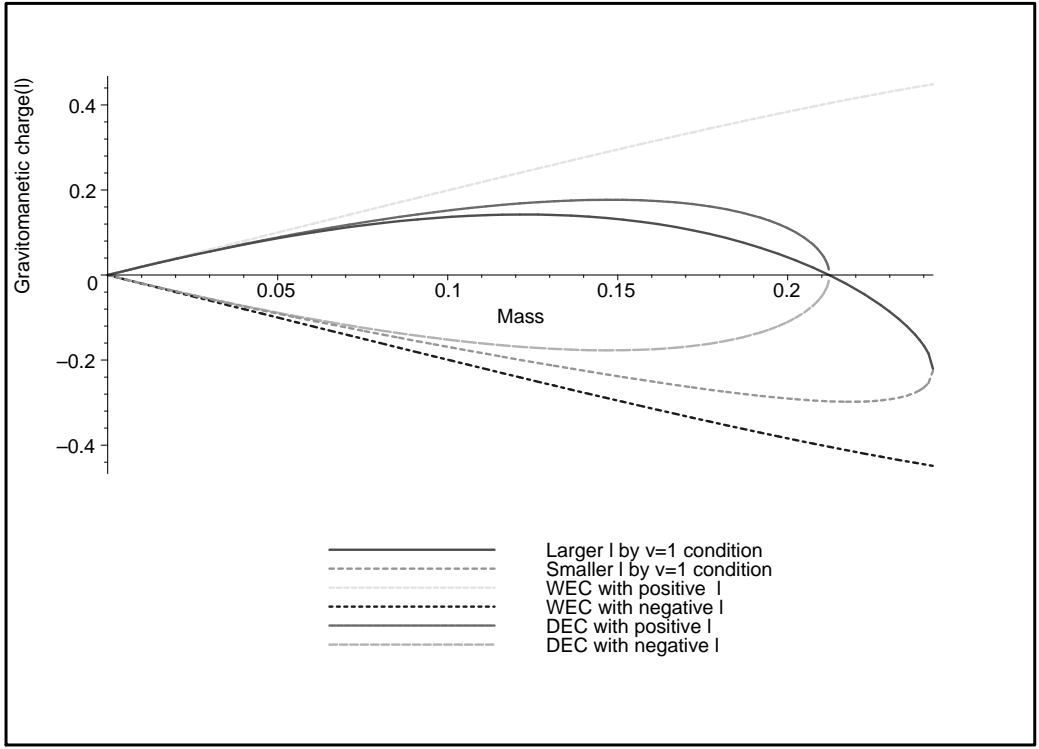}
\caption{Dominant and weak energy conditions are shown along with the velocity condition all at the rim of the disk. It is notable that there are regions for which the DEC is not satisfied but the velocity condition holds and vice versa.}
\end{center}
\end{figure}
%%%%%%%%%%%%%%%%%%%%%%%%%%%%%%%%%%%%%%%%%
\subsection{Gravitational redshift}
Using the usual formula for the gravitational redshift in an asymptotically
flat spacetime, for the MM-NUT disk space (which is asymptotically flat) we have,
$$1+z =\sqrt{{g_{tt}(\infty)\over g_{tt}(\rho)}} = 
{1\over \sqrt{1+{l^2\over 4M^2}}}\left( \exp({3\pi M\over 4}(\rho^2-2)) 
+ {l^2\over 4M^2}exp(-{3\pi M\over 4}(\rho^2 - 2))\right)^{{1\over 2}}.\eqno(74)$$
It is seen that the largest redshift occurs for an emitter at the centre of the disk ;
$$1+z_{max} =\sqrt{{g_{tt}(\infty)\over g_{tt}(0)}} = 
{1\over \sqrt{1+{l^2\over 4M^2}}}\left( \exp({3\pi M\over 2}) 
+ {l^2\over 4M^2}exp(-{3\pi M\over 2})\right)^{{1\over 2}},\eqno(75)$$
which compared to the Morgan-Morgan disk has the extra $l$-dependent term in the right 
hand side. In Fig. 5 redshift of photons emitted from the center of the disk in terms of 
NUT parameter for two different disk masses are given. We note from the above discussions that 
in general for each value of $M$ there is a range of allowed values of $l$ and indeed in Fig. 5
the masses and their corresponding NUT parameters are chosen such that  
all are physically acceptable.
\begin{figure}
\begin{center}
\includegraphics[angle =-90,scale=0.6]{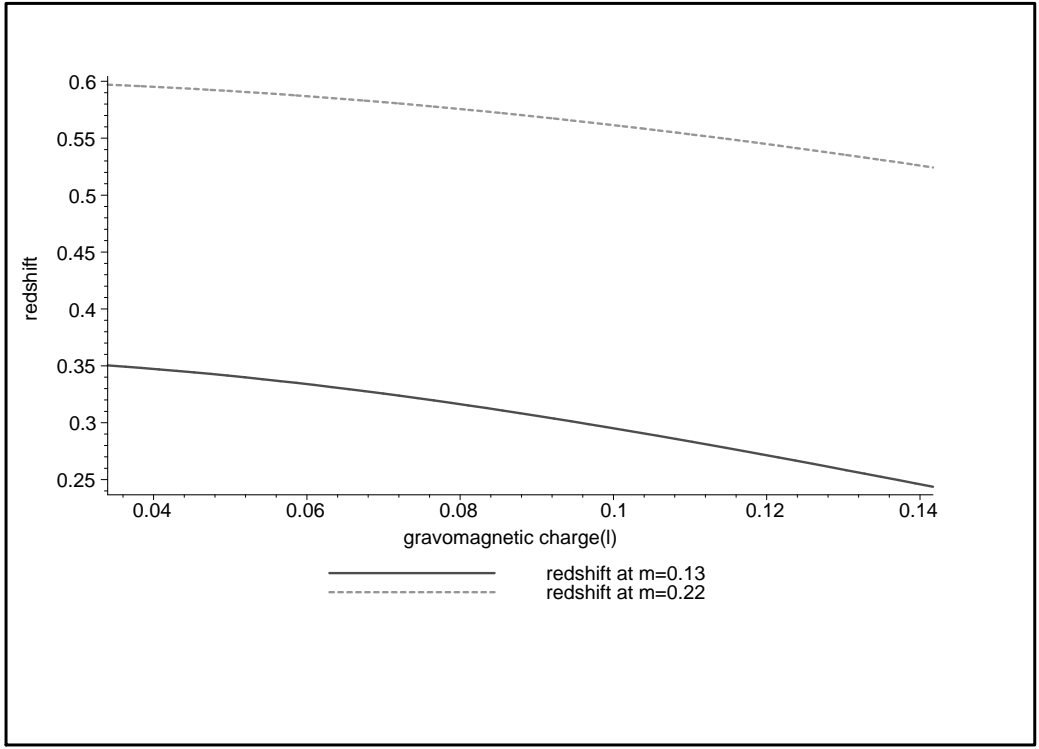}
\caption{Redshift of photons emitted from the center of the disk as a function of $l$ for two different values of disk mass.}
\end{center}
\end{figure}
%%%%%%%%%%%%%%%%%%%%%%%%%%%%%%%%%%%
\section{Summary and discussion}
In this article using the Ehlers transformation as a tool for
finding a stationary spacetime from a static spacetime and the
$1+3$ approach to spacetime decomposition which yields a physical
interpretation of the stationary spacetime in terms of
the concepts introduced under the general name of
gravitomagnetism, we found the stationary spacetime of a finite
disk dual to the static spacetime of a non-rotating disk 
(without radial pressure) introduced by Morgan and Morgan [2]. 
The exact form of this metric, which we called MM-NUT disk space,
is given in oblate ellipsoidal coordinates in (45). 
The axially symmetric nature of the spacetime and the form of the 
gravitomagnetic potential (44), infers a gravitomagnetic field with 
components along the polar directions $\rho$ and $\theta$.
The fact that we find the extra parameter of the spacetime in the
combination $\alpha M$ suggests that the parameter $\alpha$ could
be interpreted as the gravitomagnetic monopole charge per unit mass.
It is shown that the MM-NUT stationary disk admits, through the 
introduction of the NUT factor, a higher maximum disk mass than the
static Morgan-Morgan disk for a negative value of $l$. 
It is found that this spacetime is singular either in the positive or
the negative half axis, leading to the existence of closed timelike 
curves (CTCs) in the singular sector.
We also showed that MM-NUT disk has stable particle orbits and
satisfies the strong energy condition. The weak energy 
condition on the other hand holds as long as the values 
of mass and NUT parameter are such that the particle velocities do not 
exceed that of light. Interesting enough, it is shown that the dominant energy condition
could be violated for some values of $l$ and $M$ which respect the velocity 
barrier condition and vice versa.
Also as a by-product and as an example of the application of the 
Ehlers transformation, we showed that the seed metric for both NUT 
and pure NUT ($M=0$) spaces is the Schwarzschild metric.
%%%%%%%%%%%%%%%%%%%%%%%%%%
\section*{Acknowledgement}
The authors would like to thank university of Tehran for supporting
this project under the grants provided by the research council. M. N-Z 
thanks D. Lynden-Bell for useful discussions and comments.
%%%%%%%%%%%%%%%%%%
%\pagebreak


\begin{references}
\bibitem{1} J. Ehlers, (1962) Colloques Internationaux CNRS (Les theories relativistes
de la gravitation) {\bf 91}, 275.
\bibitem{2} T. Morgan, and L. Morgan, Physical Review, Vol. 183, No. 5, 1969.
\bibitem{3} E. T. Newman, L. Tamburino  and T. Unti , J. Math. Phys., {\bf 4}, 915, (1963).
\bibitem{4} H. E. J. Curzon, Proc. London Math. Soc. {\bf 23},477, (1924).
\bibitem{5} J. Chazy, Bull. Soc. Math. France, {\bf 52}, 17 (1924).
\bibitem{6} R. De Felice and J. Clarke, {\it Relativity on curved manifolds}, Cambridge 
University Press, 1990.
\bibitem{7} L. D. Landau and E. M. Lifishitz, {\it Classical theory of
fields}, 4th edn., Pergamon press, Oxford, 1975.
\bibitem{8} D. Lynden-Bell  and M. Nouri-Zonoz, Rev. of Modern Phys., Volume 70, No.2, 427 (1998).
\bibitem{9} M. Nouri-Zonoz and A.R. Tavanfar, Journal of High Energy Physics 302(2003)059.
\bibitem{10} R. T. Jantzen, P. Carini and D. Bini, Ann. Phys. NY {\bf 215}, 1, (1992).
\bibitem{11} H. Stephani, D. Kramer, M. A. H. McCallum, C. Hoenselares
and E. Herlt, {\it Exact solution of Einstein's Field equations}, Cambridge University Press, 2003.
\bibitem{12} J. Bicak, Lect. Notes Phys., {\bf 540}, 1-126, (2000).
\bibitem{13} M. Halilsoy, J. Math. Phys., {\bf 33}, 4225 (1992).
\bibitem{14} M. Halilsoy, and O. Gurtug, Nuovo Cim. B, {\bf 109}, 963 (1994).
\bibitem{15} M. Nouri-Zonoz, Class. Quantum Grav. {\bf 14} (1997) 3123-3129.
\bibitem{16} Letelier, P. S. and Oliviera, S. R, Class. Quantum Grav., {\bf 15} (1998) 421.
\bibitem{17} J. Bicak, D. Lynden-Bell and J. Katz, Phys. Rev. D {\bf 47}, 4334 (1993).
\bibitem{18} J. Bicak, D. Lynden-Bell and C. Pichon, Mon. Not. R. Astron. Soc. 265, 126 (1993).
\bibitem{19} J. P. S. Lemos and P.S. Letelier, Phys. Rev. D {\bf 49}, 5135-5143 (1994). 
\bibitem{20} D. Lynden-Bell and S. Pineault, Mon. Not. R. Astron. Soc. 185, 679 (1978).
\bibitem{21} J. P. S. Lemos, Class. Quantum Grav. {\bf 6} (1989) 1219 .
\bibitem{22} S. Chandrasekhar, {\it The mathematical theory of black holes}, 
Oxford University Press, 1983.
\bibitem{23} G. A. Gonzalez, and P. S. Letelier, Phys. Rev. D {\bf 62} (2000) 064025, gr-qc/0006002.
\bibitem{24} P. S. Letelier, Phys. Rev. D {\bf 68} (2003) 104002, gr-qc/0309033.



\end{references}
\end{document}